\documentclass[12pt]{article}
\pdfoutput = 1
\usepackage{graphicx}
\usepackage{adjustbox}
\usepackage{fancybox}
\usepackage{amssymb}
\usepackage{amsthm}
\usepackage{amsfonts,amsmath}
\usepackage{tikz} 
\usetikzlibrary{fit,positioning}
\usepackage[section]{placeins}
\usepackage{lineno}
\usepackage{caption}
\usepackage{subcaption}
\usepackage{float}
\usepackage{adjustbox}
\usepackage{multirow}
\usepackage{authblk}
\usepackage{nameref,hyperref}

\title{Common patterns between dengue cases, climate, and local environmental variables in Costa Rica: A Wavelet Approach.}

\author[1]{Yury E. García}
\author[2]{Shu-Wei Chou-Chen}
\author[4]{Luis A. Barboza}
\author[1]{Maria L. Daza–Torres}
\author[1]{J. Cricelio Montesinos-L\'opez}
\author[3]{Paola V\'asquez}
\author[4]{Juan G. Calvo}
\author[1]{Miriam Nu\~no}
\author[4]{Fabio Sanchez}

\affil[1]{Department of Public Health Sciences, 
University of California Davis, CA, USA.}

\affil[2]{Centro de Investigaci\'on en Matem\'atica Pura y Aplicada - Escuela de Estadística, Universidad de Costa Rica, San Jos\'e, Costa Rica.}

\affil[3]{Centro de Investigaci\'on en Matem\'atica Pura y Aplicada, Universidad de Costa Rica, San Jos\'e, Costa Rica.}

\affil[4]{Centro de Investigaci\'on en Matem\'atica Pura y Aplicada - Escuela de Matem\'atica, Universidad de Costa Rica, San Jos\'e, Costa Rica}

\date{}

\begin{document}
\maketitle

\begin{abstract}
Throughout history, prevention and control of dengue transmission have challenged public health authorities worldwide. In the last decades, the interaction of multiple factors, such as environmental and climate variability, has influenced increments in incidence and geographical spread of the virus. In Costa Rica, a country characterized by multiple microclimates separated by short distances, dengue has been endemic since its introduction in 1993. Understanding the role of climatic and environmental factors in the seasonal and inter-annual variability of disease spread is essential to develop effective surveillance and control efforts. In this study, we conducted a wavelet time series analysis of weekly climate, local environmental variables, and dengue cases (2001-2019) from 32 cantons in Costa Rica to identify significant periods (e.g., annual, biannual) in which climate and environmental variables co-varied with dengue cases. Wavelet coherence analysis was used to characterize seasonality, multi-year outbreaks, and relative delays between the time series. Results show that dengue outbreaks occurring every 3 years in cantons located in the country's Central, North, and South Pacific regions were highly coherent with the Oceanic Ni\~no 3.4 and the Tropical North Caribbean Index (TNA). Dengue cases were in phase with  El Ni\~no 3.4 and TNA, with El Ni\~no 3.4 ahead of dengue cases by roughly nine months and TNA ahead by less than three months. Annual dengue outbreaks were coherent with local environmental variables (NDWI, EVI, Evapotranspiration, and Precipitation) in most cantons except those located in the Central, South Pacific, and South Caribbean regions of the country. The local environmental variables were in phase with dengue cases and were ahead by around three months.
\end{abstract}

\section*{Introduction}
Dengue is a mosquito-borne viral infection caused by four antigenically distinct dengue virus serotypes (DENV1-4). The transmission to humans occurs by the bite of an infected female mosquito. The \textit{Aedes aegypti} is the primary vector, more common in rural areas, and \textit{Aedes albopictus} is a secondary vector, considered one of the 100 worst invasive species in the world~\cite{outammassine2022global}, currently present on all continents except Antarctica~\cite{medlock2006analysis,romi2006cold}. Dengue is a flu-like illness that affects individuals of all ages, causing significant health, economic, and social burdens on a population~\cite{shepard2011economic}. The clinical profile of patients can range from asymptomatic infection to severe cases.

In recent years, the complex interaction of biological, socioeconomic, environmental, and climatic factors has facilitated the rapid emergence of this viral infection throughout the world, becoming endemic and a relevant public health problem in more than 100 countries~\cite{zeng2021global}. In the last decades, the number of dengue cases reported to the World Health Organization (WHO) has increased from 505,430 cases in 2000 to more than 4.2 million in 2019~\cite{dengueanWHO,zeng2021global}.

Seasonal case patterns and vector abundance suggest that dengue transmission is sensitive to climatic and environmental factors~\cite{morin2013climate,kolivras2010changes}. Climatic conditions can alter spatial and temporal dynamics of vector ecology, potentially increasing vector ranges, lengthening the duration of vector activity, and increasing the mosquito’s infectious period~\cite{morin2013climate}. Precipitation provides habitats for the aquatic stages of the mosquito life cycle and strongly influences vector distribution~\cite{kolivras2010changes,tun477effects}. On the other hand, water temperature plays a significant role in mosquito reproduction since it directly affects survival at all stages of its life cycle~\cite{ebi2016dengue}. Temperature increases are associated with a faster rate of viral replication within the vector and a shorter extrinsic incubation period~\cite{morin2013climate}. Furthermore, higher humidity is associated with an increase in \textit{Ae. aegypti} feeding activity, which enhances the spread of the diseases.

The complexity of dengue transmission has driven many studies to assess its correlation with meteorological and ecological variables~\cite{prabodanie2020coherence,ehelepola2015study, cuong2016quantifying,thai2010dengue,johansson2009multiyear,cazelles2005nonstationary,jury2008climate}. Most of these works evaluated the effects and correlation between dengue cases and climate variables~\cite{nitatpattana2007potential,nakhapakorn2020assessment} and seasonal vegetation dynamics, which may also influence the biology of the vector populations at relatively local scales ~\cite{chaves2021modeling,mudele2021modeling,troyo2009urban}. Barrera et al.~\cite{barrera2006ecological} suggested that dense vegetation can promote \textit{Ae. aegypti} pupal productivity by contributing organic material to the habitat and influencing water temperature and evaporation by creating shades. 

This study focuses on Costa Rica, where, in 1961, successfully eradicated the mosquito \textit{Ae. aegypti} after intense prevention and control campaigns. In 1970, the lack of continuity in active surveillance caused the mosquito to be found again on the Pacific coast. By September 1993, health officials reported the first dengue case. The patterns and periodicity of transmission are different across the country. Despite the coastal regions being the most affected areas, trends observed over the years show variations in transmission peaks in all affected areas, challenging public health authorities to allocate and optimize available resources.

This work aims to describe the incidence patterns of dengue in 32 different cantons (of interest to the Ministry of Health) of Costa Rica and its correlation with climatic and local environmental variables using wavelet coherence and wavelet cluster analysis. Wavelet transform decomposes a time series in both the time and frequency domains, revealing how different periods change over time into non-stationary signals~\cite{cazelles2007time,cazelles2008wavelet}. Furthermore, it allows conclusions to be drawn about the synchronicity of the series in specific periods~\cite{cazelles2005nonstationary}. Wavelets have been used to study time series with different purposes: to evaluate the main characteristics of non-stationary time series~\cite{talagala2015wavelet, cazelles2005nonstationary, nagao2008decreases,johansson2009multiyear,cuong2013spatiotemporal}, to analyze spatial patterns~\cite{cazelles2005nonstationary,thai2010dengue}, to study the relationship between population and environmental time series; finding the phase and/or synchrony patterns~\cite{johansson2009multiyear,cazelles2005nonstationary}, and to study multiple time series~\cite{d2012wavelets,aghabozorgi2015time}.

In this study, we describe and compare the relationships between El Ni\~no-Southern Oscillation (ENSO), Tropical North Caribbean Index (TNA), Normalized Difference Water Index (NDWI), Enhanced Vegetation Index (EVI), Evapotranspiration (ET), precipitation, and dengue cases in 32 cantons in Costa Rica using a wavelet approach.

\section*{Materials and methods}
\label{Methods}

\subsection*{Study Area}

\begin{figure}[!ht]
\includegraphics[scale=0.55]{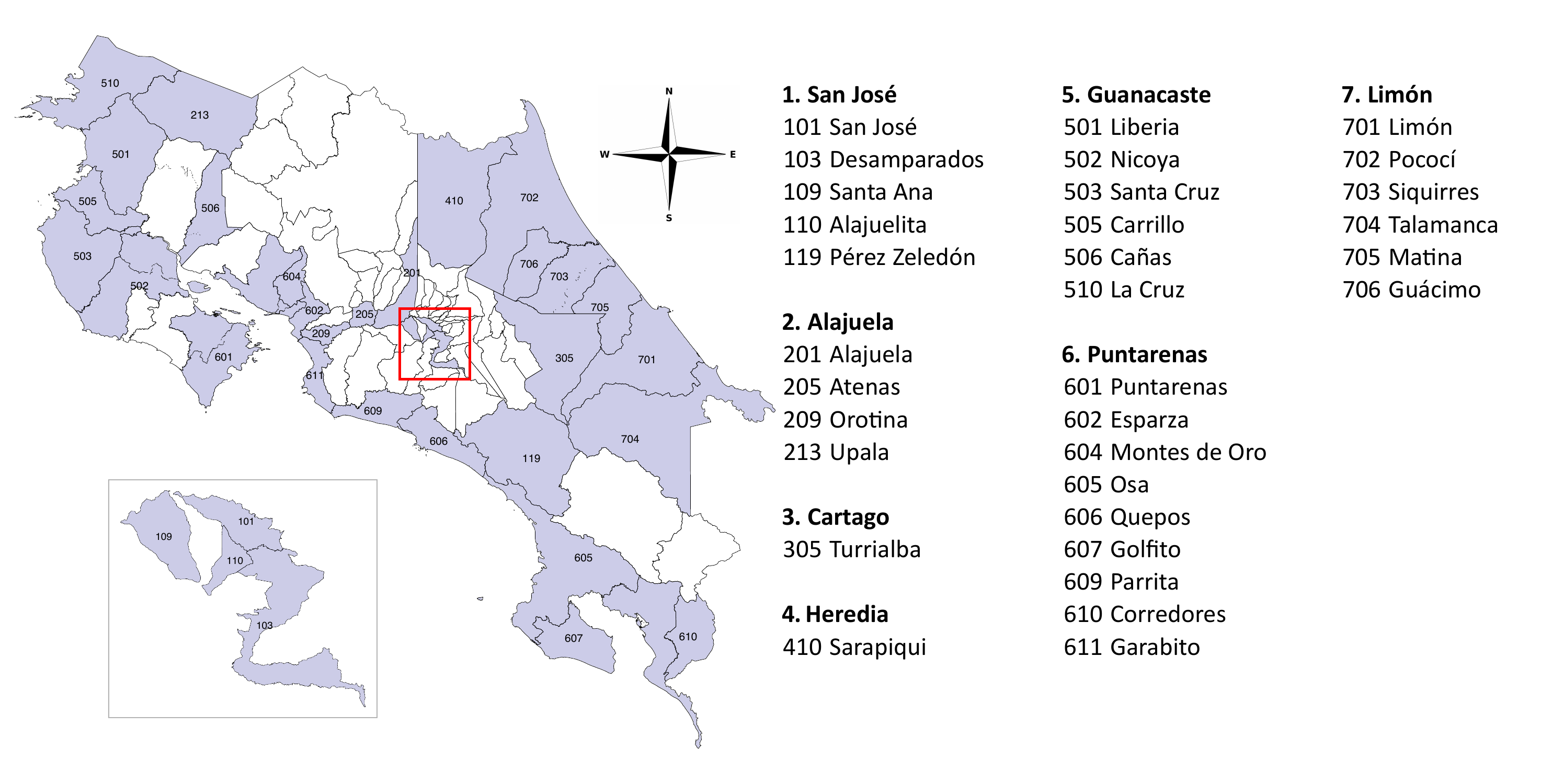}
\caption{{\bf Costa Rica Study Area.}
The colored cantons correspond to the 32 cantons highlighted as being of interest to the Minister of Health of Costa Rica due to the prevalence of dengue.}
\label{fig:MapCR}
\end{figure}

Costa Rica is a Central American country bordering Nicaragua to the North, the Caribbean Sea to the northeast, Panama to the southeast, and the Pacific Ocean to the Southwest. It has a population of around five million in a land area of 51,060 $km^2$. Almost half of the population is concentrated in the Great Metropolitan Area, which includes the capital, San Jos\'e. Costa Rica is divided into seven provinces: San Jos\'e, Alajuela, Heredia, Cartago, Guanacaste, Puntarenas, and Lim\'on. These provinces are further divided into different cantons (83 in total). Due to the high number of dengue cases, we focused this analysis on 32 cantons reported by the local public health authorities as areas of interest (Fig~\ref{fig:MapCR}).

The country has a tropical climate with various microclimates. On the Pacific coast and the Central region, there are a dry season from December to April and a rainy season from May to November, during which rainfall is abundant. In contrast, in the eastern plains and coasts (but also in the southern part of the Pacific coast), the climate is equatorial, with abundant rainfall throughout the year. 

In Costa Rica, the success in the fight against \textit{Ae. aegypti} in the 1950s led to the declaration of the country as free of the vector by 1961. However, the lack of continuity in active surveillance caused, around 1971, the detection of positive places for the presence of the mosquito. After that time, a new eradication campaign was established. However, by 1992, the vector was already in almost all national territories. By September 1993, the first dengue cases were reported on the Pacific coast. Since then, the disease has exhibited endo-epidemic transmission with the circulation of the four dengue virus serotypes. According to data from the Ministry of Health, from 1993 to 2021, the country has notified a total of 398.546 cases. During these years, transmission has been characterized by epidemic peaks occurring every 2 to 5 years, with 2013 being the year with the highest number of cases (49,993 cases) reported by health centers around the country, followed by 2005 (37,798 cases) and 2010 (31,484 cases)~\cite{SitioWeb5}. The year 2007 reported the highest number of severe dengue cases for 318 patients, which represented 1.19\% of the total cases reported for that period. Despite the coastal regions being the most affected areas, trends observed over the years show variations in transmission peaks in all affected areas (Fig S1).  

\subsection*{Data}
\subsubsection*{Dengue cases}

The Ministry of Health of Costa Rica provided weekly dengue case records for all cantons. For this analysis, the data were aggregated monthly, square-root transformed, and standardized due to their asymmetry~\cite{cazelles2005nonstationary}. 

\subsubsection*{Climate variables}
Anomalies in the Caribbean and Pacific Ocean were considered throughout the Tropical North Caribbean Index (TNA)~\cite{TNA} and El Ni\~no-Southern Oscillation (ENSO) for region 3.4~\cite{CDG}, respectively~\cite{hanley2003quantitative,TNA,wu2002tropical}. 

The TNA SST anomaly index indicates the surface temperatures in the eastern tropical North Caribbean Ocean. El Ni\~no (positive phase of ENSO) occurs, on average, every 3-7 years with episodes typically lasting 9-12 months~\cite{ClimateP23}, and is characterized by sea surface temperature (SST) above the mean in region 3.4 of the equatorial Pacific~\cite{ClimateP23}. ENSO is the Earth’s strongest inter-annual climate cycle and is the leading cause of climatic variability in Northeastern South America~\cite{cai2020climate}. ENSO data was obtained from the Climate Prediction Center (CPC) of the United States National Oceanographic and Atmospheric Administration (NOAA)~\cite{CDG}. 

\subsubsection*{Environmental variables}
We considered four environmental variables that provide local information on vegetation, precipitation, and evapotranspiration. The \textit{Normalized Difference Water Index} (NDWI) describes changes in the liquid water content of leaves, providing a proxy for water stress in vegetation~\cite{parselia2019satellite,gao1996ndwi,NDWI}. The \textit{Enhanced Vegetation Index} (EVI) provides spatial and temporal information about vegetation. It can be used to quantify vegetation greenness~\cite{GisGeography}. \textit{Evapotranspiration} (ET) can be used to calculate regional water and energy balance and soil water status; hence, it provides key information for water resource management~\cite{MODISET}. Finally, precipitation index corresponds to rainfall estimates and was obtained from CHIRPS (Climate Hazards Group InfraRed Precipitation with Station)~\cite{CHIRPSR}. All data was standardized to provide consistency and allow for comparisons across different time series and corresponding results.

\subsection*{Wavelets analysis}

To understand the time-frequency variability of dengue, climate, and environmental variables, we conducted a wavelet time series analysis over a time period of nineteen years (2001-2019). Wavelet time series analyses are ideal for noisy, non-stationary data, such as dengue case data, demonstrating strong seasonality and inter-annual variability (yearly changes)~\cite{meyers1993introduction,cazelles2007time}. In the following, we present the critical details of our analyses. For further depth, Torrence and Compo \cite{torrence1998practical} provide a comprehensive presentation of wavelet time series analysis, and Cazellez et al.~\cite{cazelles2008wavelet,cazelles2007time,cazelles2014wavelet} offer a perspective of the use of these techniques in ecological and epidemic applications.

\subsubsection*{Wavelet power spectra}

The wavelet analysis is based on a wavelet transform defined as:

\begin{equation}
W_x(s,\tau)=\dfrac{1}{\sqrt{s}}\int_{-\infty}^{\infty}x(t)\Psi^*\left(\dfrac{t-\tau}{s}\right)dt  
=\int_{-\infty}^{\infty}x(t)\Psi^{*}_{s,\tau}(t)dt,
\end{equation}    
where $*$ denotes the complex conjugate form and $\Psi_{s,\tau}(t)$ represent a family of functions derived from a single function called the ``mother wavelet''. The signal is decomposed in these functions which can be expressed in terms of two parameters, one for the time position $\tau$, and the other for the scale of the wavelets $s$, given by

\begin{equation}
\Psi_{s,\tau}(t) = \dfrac{1}{\sqrt{s}}\Psi\left(\dfrac{t-\tau}{s}\right) .
\end{equation}

For this analysis, we use the R-packages \textit{WaveletComp}~\cite{rosch2016waveletcomp} that analyzes the frequency structure of uni- and bivariate time series using
the Morlet mother wavelet~\cite{cazelles2007time, cazelles2008wavelet}
\begin{equation}
\Psi(t) = \pi^{\frac{-1}{4}}e^{i\omega t}e^{\frac{-t^2}{2}}.
\end{equation}

This leads to a continuous complex-valued wavelet transform that can be separated into its real and imaginary parts, providing information on both local amplitude and instantaneous phase of any periodic process across time — a prerequisite for investigating coherency between two-time series~\cite{rosch2016waveletcomp}.

\subsubsection*{Analyzing two time series}

Cross-wavelet and wavelet coherence allowed us to compare two time series, such as climate and dengue, and to identify synchronous periods or signals. The \textit{cross-wavelet transform} of two-time series ($x_t$) and ($y_t$), with respective wavelet transforms $W_x$, and $W_y$ decomposes the Fourier co- and quadrature-spectra in the time-frequency (or time-scale) domain. The concepts of cross-wavelet analysis provide a tool for (i) comparing the frequency of two-time series, (ii) concluding the series’ synchronicity at specific periods and across certain ranges of time~\cite{rosch2016waveletcomp}. The cross-wavelet transform of two time series ($x_t$) and ($y_t$) is given by

$$W_{x,y}(\tau,s)=\dfrac{1}{s}W_x(\tau,s)W_y^*(\tau,s)$$

Its modulus can be interpreted as \textit{cross-wavelet power}.

$$Power.xy(\tau,s)=|W_{x,y}(\tau,s)|$$

In a geometric sense, the \textit{cross-wavelet transform} is the analog of the covariance. However, it lends itself to certain limitations for interpretation concerning the degree of association of the two series that can be remedied by coherence.

\subsubsection*{Wavelet coherence}

\textit{Wavelet Coherence} is a normalized measure of dependence for which it is possible to construct confidence intervals, commonly considered more interpretable than the cross-wavelet transform~\cite{rosch2016waveletcomp}. Fourier coherency measures the cross-correlation between two time series as a function of frequency; an analogous concept in wavelet theory is the notion of wavelet coherency. In a geometric sense, coherency is the analog of classical correlation. Consequently, in analogy with the notion of Fourier coherence and the coefficient of determination in statistics, wavelet coherence is given by squared coherence.

\begin{equation}\
R_{x,y}(s,\tau) =\dfrac{|\langle W_{x,y}(s,\tau)\rangle |^2}{|\langle W_x(s,\tau)\rangle|^2|\langle W_y(s,\tau)\rangle|^2}    
\end{equation}

The angle brackets indicate smoothing in both time and frequency, $W_x(s, \tau) $ and $W_y(s,\tau)$ are the wavelet transform of the series $x_t$ and $y_t$, respectively. The value of $R_{x,y}(s,\tau)$ ranges between 0 and 1, where 1 represents a perfect linear relationship between the time series $x_t$ and $y_t$.

The phase difference associated to the two signals, provides information about series synchronization (i.e., in phase or out of phase). The Morlet wavelet is a complex wavelet, so the phase difference can be computed in terms of the real ($\mathcal{R}$) and the imaginary ($\mathcal{I}$) part, as follows

\begin{equation}
\Phi_{x,y}(s,\tau) = \dfrac{\mathcal{I}(\langle W_{x,y}(s,\tau)\rangle)}{\mathcal{R}(\langle W_{x,y}(s,\tau)\rangle)}
\label{eq:phase}
\end{equation}

The instantaneous time lag between the time series $x_t$ and $y_t$ is also computed \cite{cazelles2005nonstationary}.

\subsubsection*{Wavelet clusters}

Clustering is partitioning a set of objects into groups (clusters) so that objects within a group are more similar to each other than objects in different groups according to some variables. Most of the clustering algorithms depend on some assumptions in order to define the subgroups present in a data set. Consequently, the resulting clustering scheme requires some evaluation regarding its validity.

Since dengue and climatic data are time-dependent and non-stationary, this paper focuses on performing clustering analysis based on vector wavelet coherence, proposed by Oygur and Unal \cite{oygur2021vector}. This methodology allows us to measure the synchronicity and co-movements between vectors of different climatic time series and dengue to perform cluster analysis based on their synchronicity for 32 cantons.

The cluster analysis based on vector wavelet coherence is performed using Wald's method \cite{Murtagh2014}. Since the clustering is done by using a dissimilarity matrix constructed by the vector wavelets coherence, we were able to compute Silhouette \cite{Rousseeuw1987}, Frey \cite{Frey1972}, McClain \cite{McClain1975}, Cindex \cite{Hubert1985} and Dunn \cite{Dunn1974} indices, by setting minimum and maximum clusters as 2 and 15 respectively, in order to determine the optimal number of clusters. However, all these indices have differing optimal numbers of clusters. We decided to combine the Cindex criterion with the expert criteria so that the results could be interpretable.

Initially, an analysis was carried out in which all the time series were considered. However, the results could have been more readily interpretable. The cluster algorithms could only establish relationships between two or three cantons, so most groups were made up of a single item. Thus, considering the nature of the data, the analysis was divided into two moments. One in which the coherence between dengue cases and climatic variables was analyzed, and the second considers the coherence between four different local environmental variables and dengue cases.

\subsubsection*{Computing environment}

All analyses were performed using the statistical package R version 2.4~\cite{team2013r}. To perform the analysis, we first compute the multiple wavelet coherence using the \textit{vwc} function in the R-packages \texttt{vectorwavelet}~\cite{oygur2021vector}. Then, the function \textit{wclust} in the R-package \texttt{biwavelet}~\cite{gouhier2013package} is used to compute the dissimilarity matrices, and \texttt{NbClust}~\cite{Charrad2014} is used to create the clusters and indices calculation. Finally, the package \texttt{WaveletComp} \textit{version 1.1}~\cite{rosch2016waveletcomp} is considered to obtain more interpretable results on the coherence between the incidence of dengue and each climatic and environmental variable. Wavelet coherence, power average, and phase difference are plotted and grouped according to the clusters obtained with the \texttt{biowavelet} package.

\newpage
\section*{Results}
\label{Results}
\subsection*{Climate variables and dengue cases}

\begin{figure}[!ht]
\centering
\subfloat[]{\includegraphics[scale=0.4]{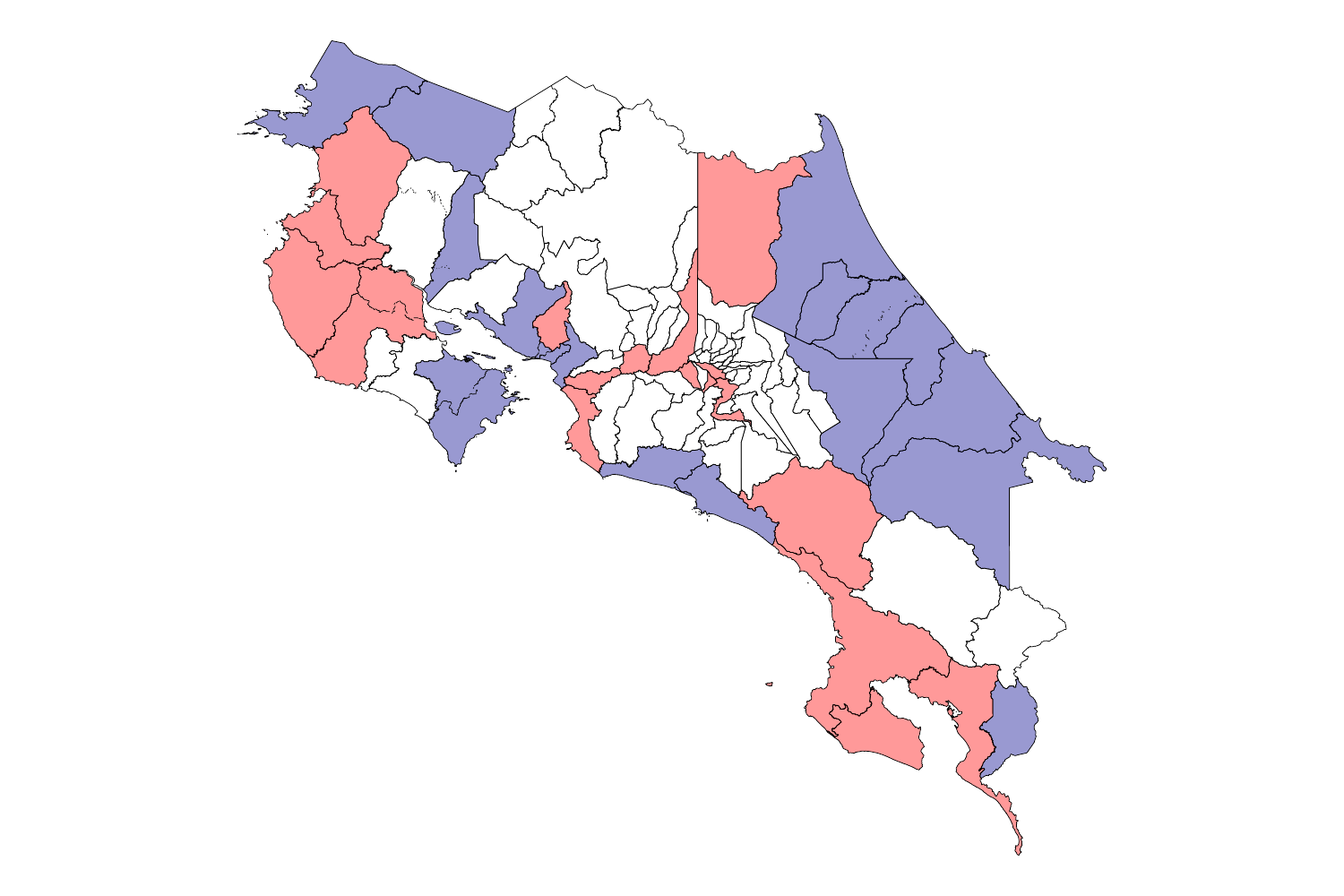}}
\subfloat[]{\includegraphics[scale=0.4]{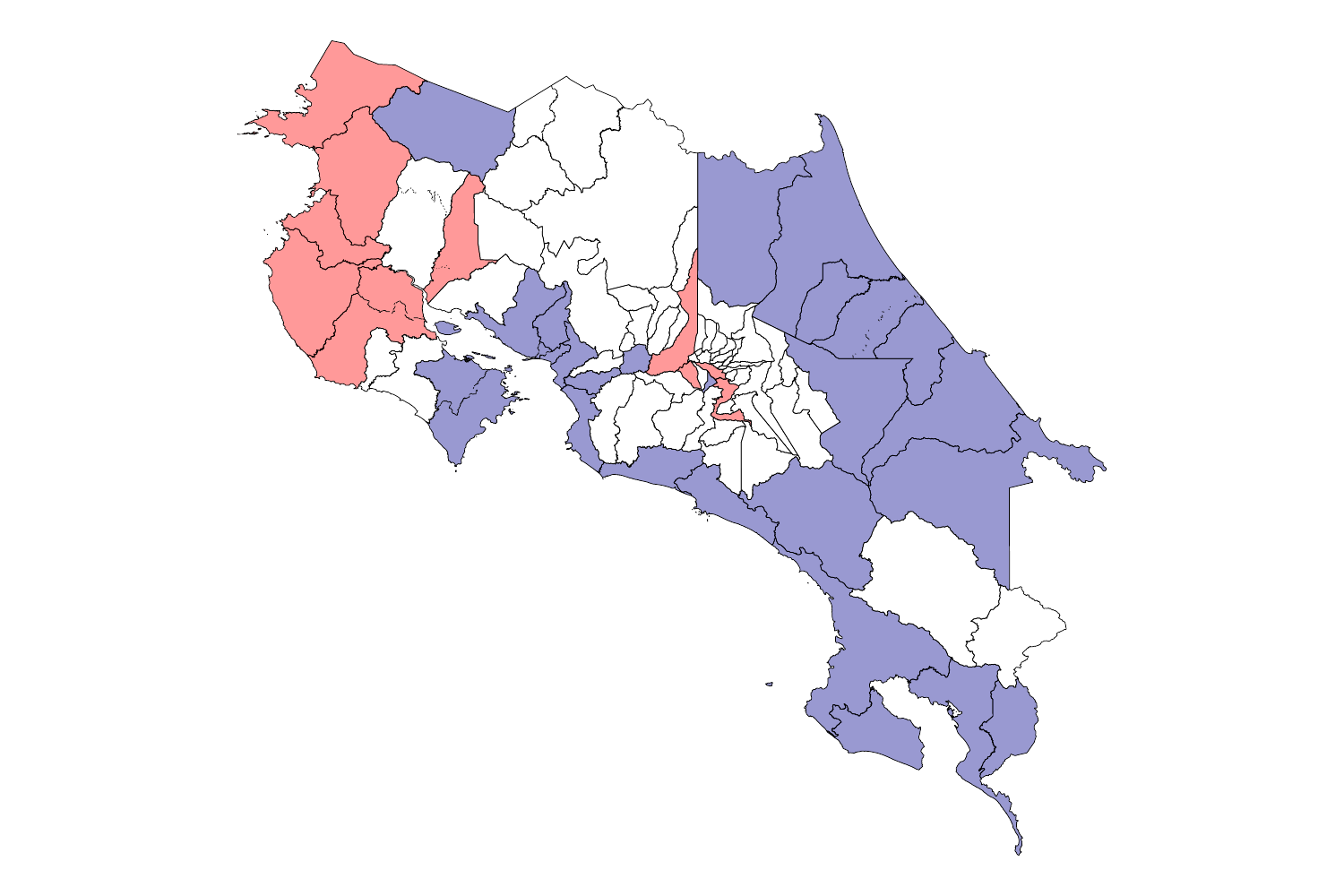}}
\caption{{\bf Cantons where dengue cases co-varied with El Ni\~no and TNA.} Fig (a) represents the cantons (red) where dengue outbreaks every 3 years showed significant coherence with El Niño and TNA. Fig (b) corresponds to those cantons (red) where the annual dengue outbreaks showed significant coherence with TNA, with TNA variable leading only by short periods. Data for cantons in white where no available for this analysis.}
\label{fig:ClimaEffect}
\end{figure}

The coherence between climatic variables (TNA and El Ni\~no 3.4) and dengue cases changes over time and geographically. A description of particular characteristics observed in the different clusters is presented below. However, some general features are common to the places where coherence was observed: 1) The 3-y dengue outbreaks were highly coherent with climate time series after 2008 in cantons located in the central and those close to the Pacific ocean (See Fig~\ref{fig:ClimaEffect}). Dengue and the climate time series were synchronized, with El Ni\~no 3.4 leading by around nine months and TNA ahead by less than three months. 2) Associations between climate and dengue cases are also observed in the 1, 1.5, and 2 yr periods, with dengue time series leading. 3) There was no significant multiyear coherence between climate variables and dengue cases in cantons mainly located in the Caribbean, and some others, such as Parrita and Quepos, in the central Pacific. 4) The cantons grouped in the clusters share behavior patterns regarding the periods in which the highest significant areas appear and the time series synchronization.

\subsubsection*{Characteristics of clusters that include climate variables (CV)}

\begin{table}[H]
\centering
\caption*{}
\label{Tab:climate}
\begin{tabular}{p{4cm}p{11.5cm}}
\textbf{Cluster 1:}& \textbf{Alajuela, Orotina, Perez Zeled\'on, San Jos\'e, Santa Ana, Sarapiqu\'i}\\
\begin{minipage}{4cm}
\includegraphics[width=1.2\textwidth]{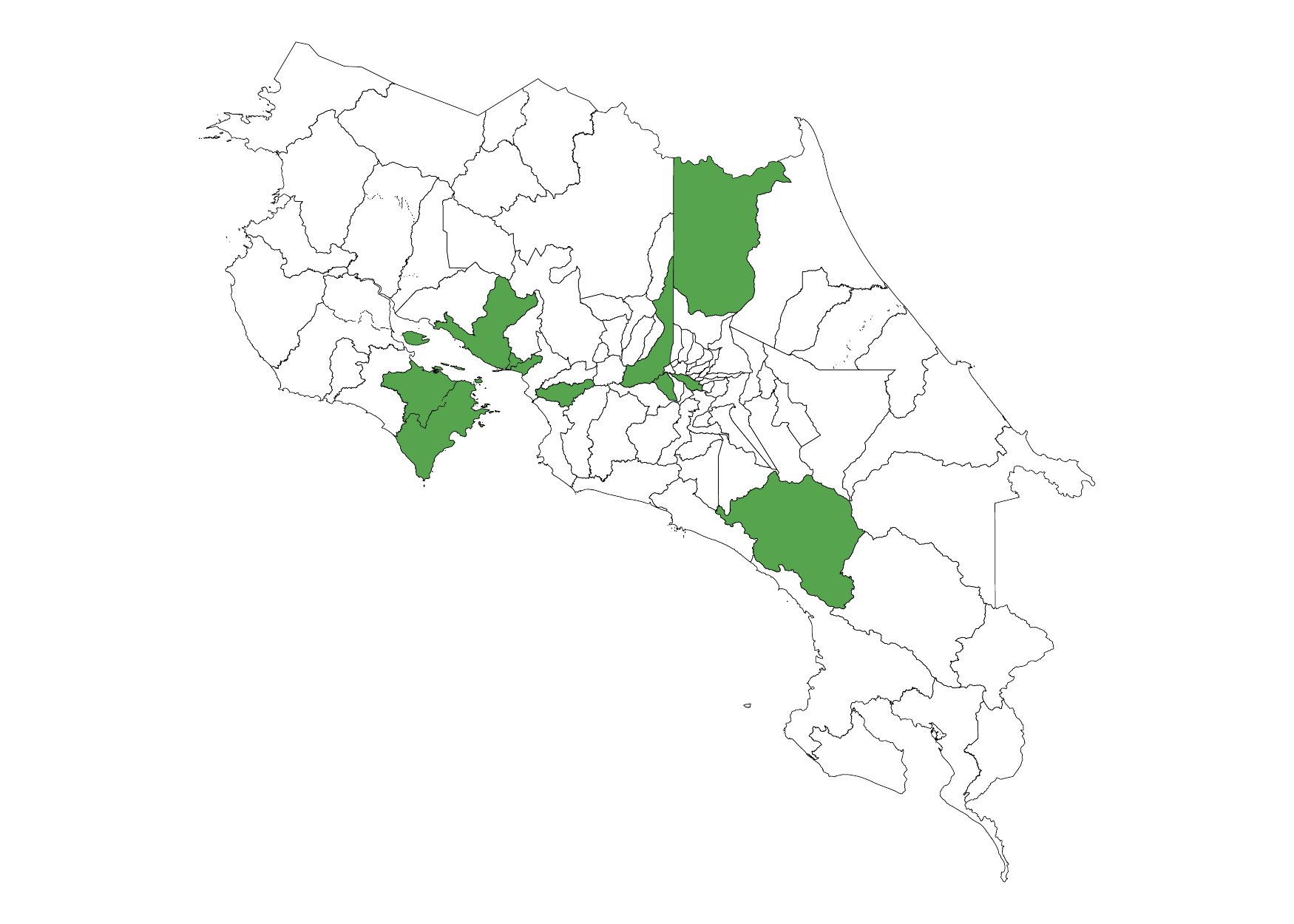}
\label{fig:ClimaCluster1}
\end{minipage}
&
\begin{minipage}{11.5cm}
\vspace{0.2cm}
Wavelet coherence shows an area of joint significance in the 3-yr period after 2010 for both TNA and El Ni\~no 3.4 with dengue cases. Except for Puntarenas and Alajuela, the coherence with TNA is mainly in the period of 1-yr and needs to be identifiable between El Ni\~no 3.4 and Puntarenas. The horizontal arrows in period 3 pointing to the right indicate that the two series TNA and dengue cases are in phase with vanishing phase differences from 2011 to 2017. For El Ni\~no 3.4 and dengue cases, the time series are also in phase after 2010, with El Ni\~no 3.4 leading by roughly 9 months.\\
\end{minipage}\\
\end{tabular}
\end{table}

\begin{table}[H]
\centering
\begin{tabular}{p{4cm}p{11.5cm}}
\textbf{Cluster 2:}& \textbf{Alajuelita, Ca\~nas, Esparza, La Cruz, Talamanca, Upala} \\
\begin{minipage}{4cm}
\includegraphics[width=1.2\textwidth]{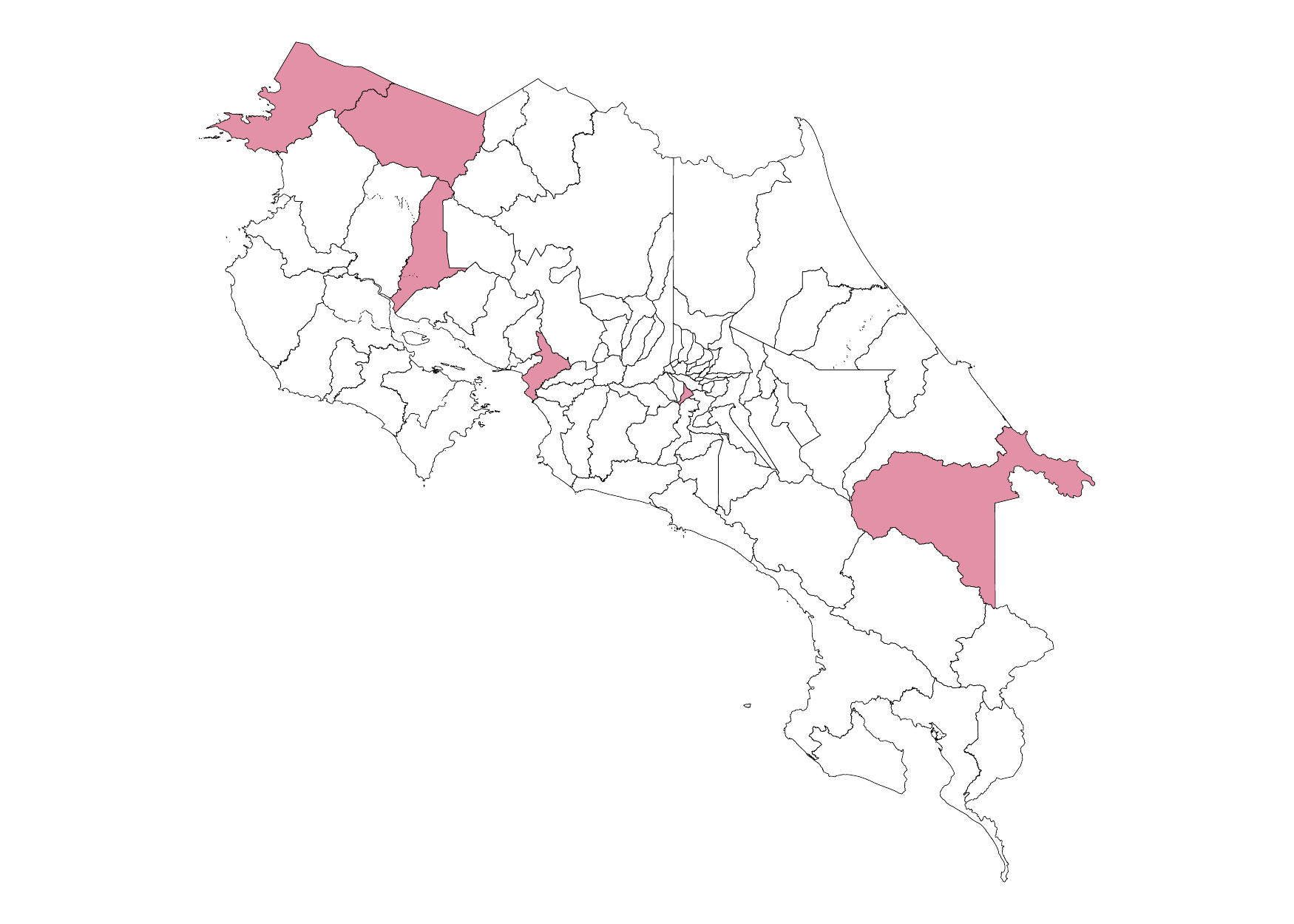}
\end{minipage}
&
\begin{minipage}{11.5cm}
\vspace{0.2cm}
Dengue incidence showed significant coherence with TNA in the 1-yr period, mainly after 2010 for all cantons except in Alajuelita, where the coherence is around the 3-yr period from 2010-2018. The time series are in phase, but the leading time series changes. In Alajuelita, Ca\~nas, and La Cruz, the TNA time series is the leading one by less than 2 months, contrary to Esparza, Talamanca, and Upala. The wavelet coherence for El Ni\~no 3.4 and dengue cases show a significant coherence in the period of 3-yr in Upala and Alajuelita, with El Ni\~no time series ahead by roughly nine months after 2007. Small areas of significance are observed for the remaining cantons in the 1 and 2-yr periods with dengue time series ahead.\\
\end{minipage}
\end{tabular}
\end{table}

\begin{table}[H]
\centering
\begin{tabular}{p{4cm}p{11.5cm}}
\textbf{Cluster 3:}&\textbf{Atenas, Desamparados, Liberia, Parrita,
Pococi, Santa Cruz.} \\
\begin{minipage}{4cm}
\includegraphics[width=1.2\textwidth]{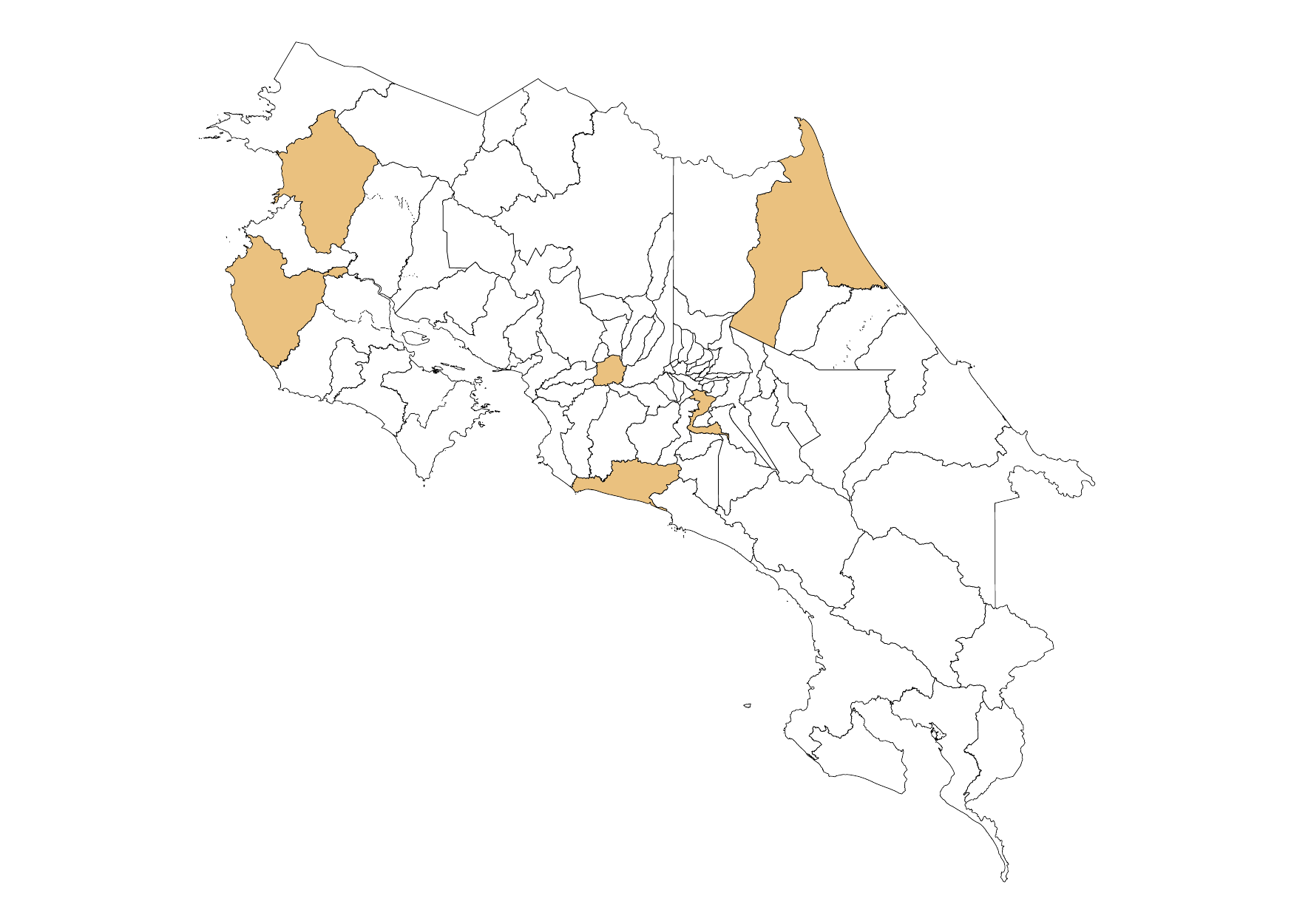}
\label{fig:ClimaCluster3}
\end{minipage}
&	
\begin{minipage}{11.5cm}
\vspace{0.2cm}
Wavelet coherence shows an area of joint significance between TNA and dengue cases in the 1-yr period, mainly between 2012 and 2017, in all cantons. However, TNA is the leading time series only in Liberia and Santa Cruz. There is also an area of joint significance in the 3-yr period between TNA and dengue cases in Atenas, Desamparados, and Santa Cruz, with the time series in-phase with vanishing phase differences between 2010-2016, except in Desamparados, where TNA was leading by a couple of weeks. The wavelet coherence for El Ni\~no 3.4 and dengue cases show an area of joint significance in the 3-yr period for Atenas, Liberia, Santa Cruz, and Desamparados. The time series are in phase after 2010 with El Ni\~no 3.4 leading by around nine months. For Parrita and Pococ\'i, there are no identifiable areas of significance.\\
\end{minipage}
\end{tabular}
\end{table}

\begin{table}[H]
\centering
\begin{tabular}{p{4cm}p{11.5cm}}
\textbf{Cluster 4:}& \textbf{Carrillo, Guacimo, Lim\'on, Nicoya, Siquirres}\\
\begin{minipage}{4cm}
\includegraphics[width=1.2\textwidth]{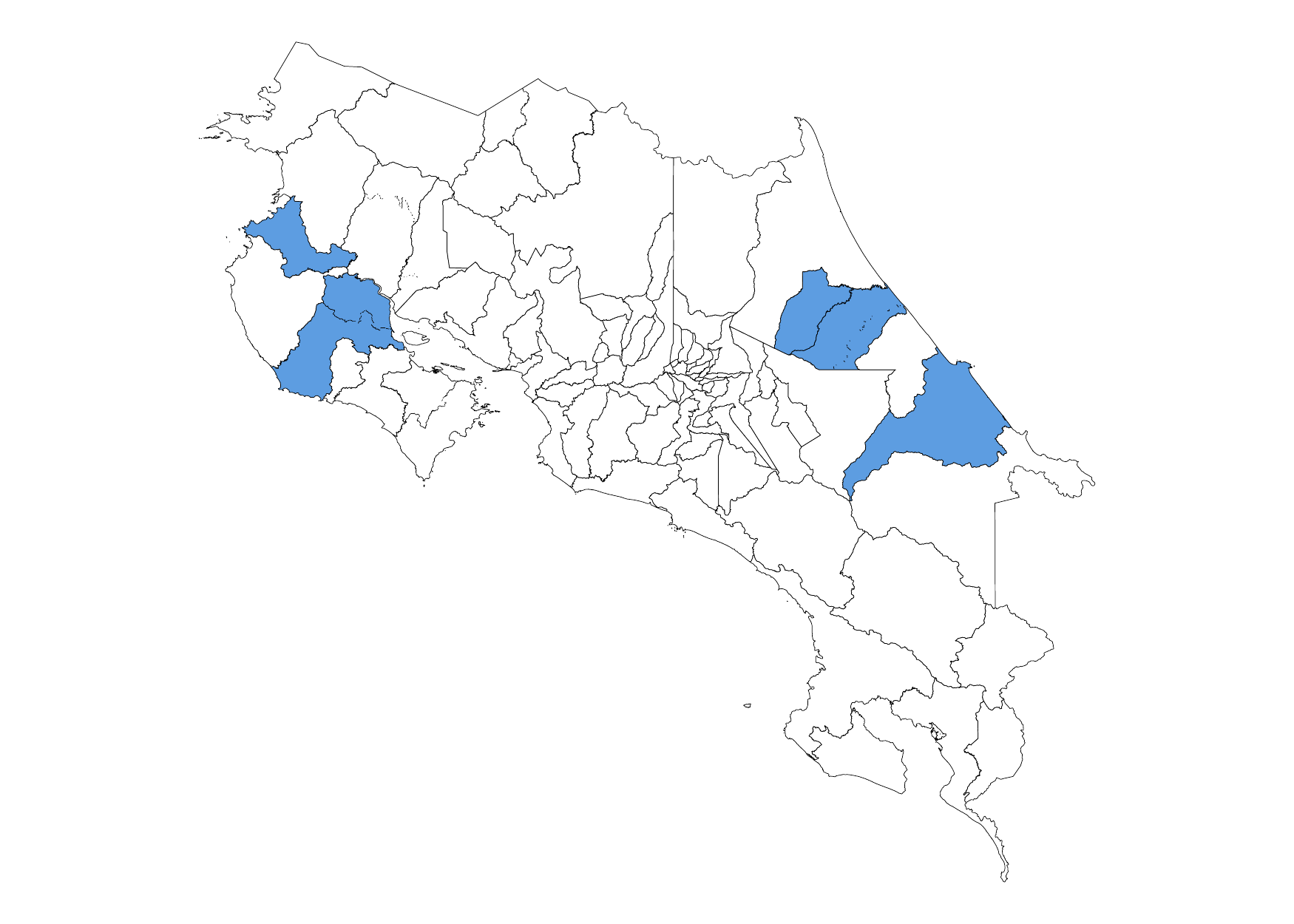}\\
\label{fig:ClimaCluster4}
\end{minipage}
&	
\begin{minipage}{11.5cm}
\vspace{0.2cm}
There are no identifiable common patterns in all cantons. Coherence for dengue cases with both TNA and El Ni\~no 3.4 shows an area of joint significance in the 3-yr period in Carrillo and Nicoya. TNA and dengue cases have vanished phase differences between 2008 and 2016 approximately in both cantons, and El Ni\~no 3.4 and dengue cases are in phase with El Ni\~no 3.4, leading by a lag of 9 months from 2008-2014. TNA and dengue cases also have an area of joint significance in the 1-yr period. The time series are in-phase after 2012, with TNA leading in Carrillo and Nicoya.\\

For the remaining cantons, the wavelet coherence shows coherence between El Ni\~no 3.4 and dengue cases in 1 and 2-yr periods, with the dengue time series ahead.
\end{minipage}\\
\end{tabular}
\end{table}

\begin{table}[H]
\centering
\begin{tabular}{p{4cm}p{11.5cm}}
\textbf{Cluster 5}& \textbf{Corredores, Golfito, Osa}\\
\begin{minipage}{4cm}
\includegraphics[width=1.2\textwidth]{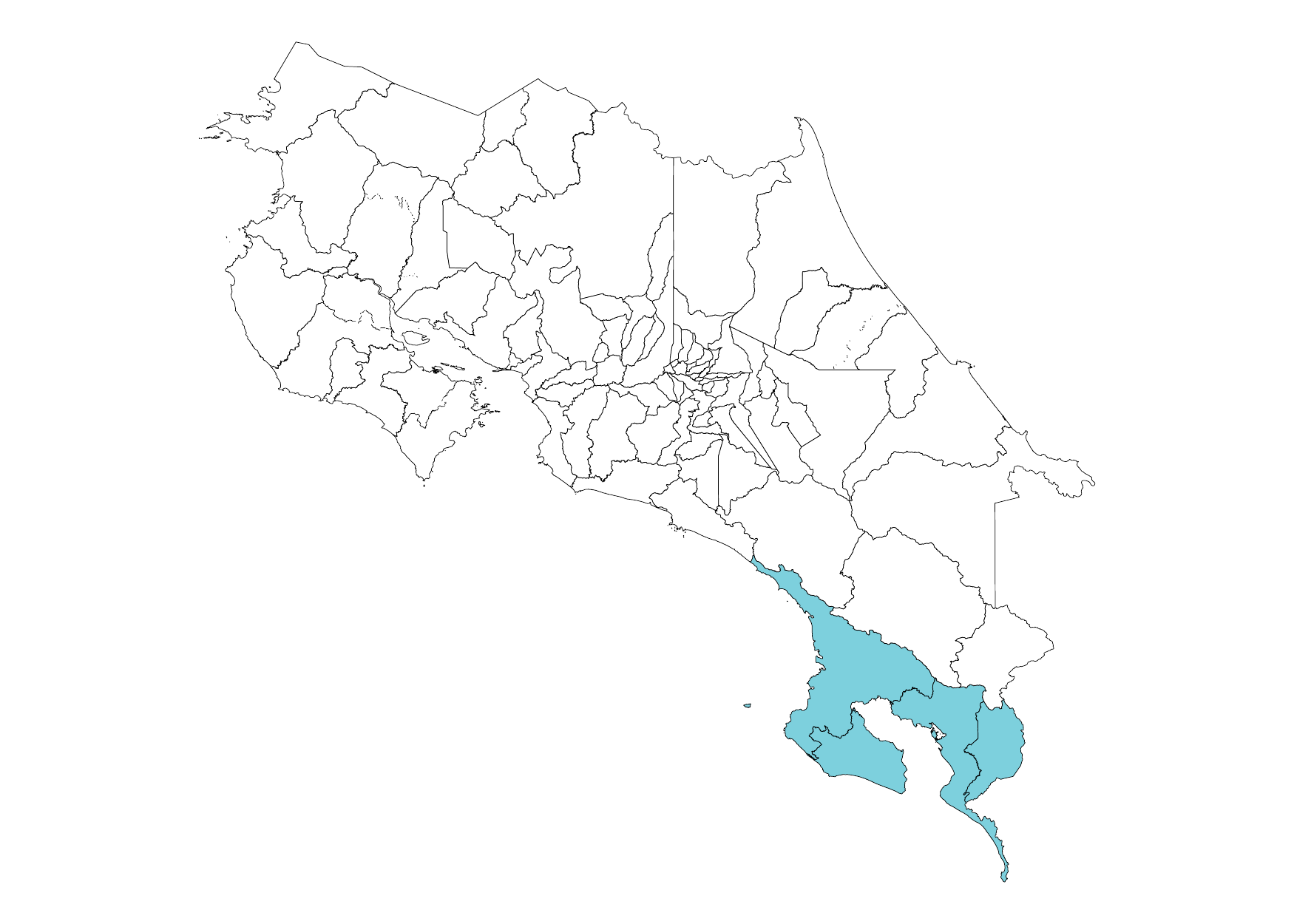}
\label{fig:ClimaCluster5}
\end{minipage}
&	
\begin{minipage}{11.5cm}
\vspace{0.2cm}
There is an area of joint significance between TNA and dengue cases in all cantons in the 1-yr and 3-yr periods where the leading time series is dengue cases. The 3-yr dengue outbreaks were highly coherent with El Ni\~no 3.4 after 2007, with El Ni\~no 3.4 leading by a lag ranging between 0 to 9 months.\\
\end{minipage}\\
\end{tabular}
\end{table}

\begin{table}[H]
\centering
\begin{tabular}{p{4cm}p{11.5cm}}
\textbf{Cluster 6}& \textbf{Garabito, Matina, Montes de Oro,
Quepos, Turrialba}\\
\label{fig:ClimaCluster6}
\begin{minipage}{4cm}
\includegraphics[width=1.2\textwidth]{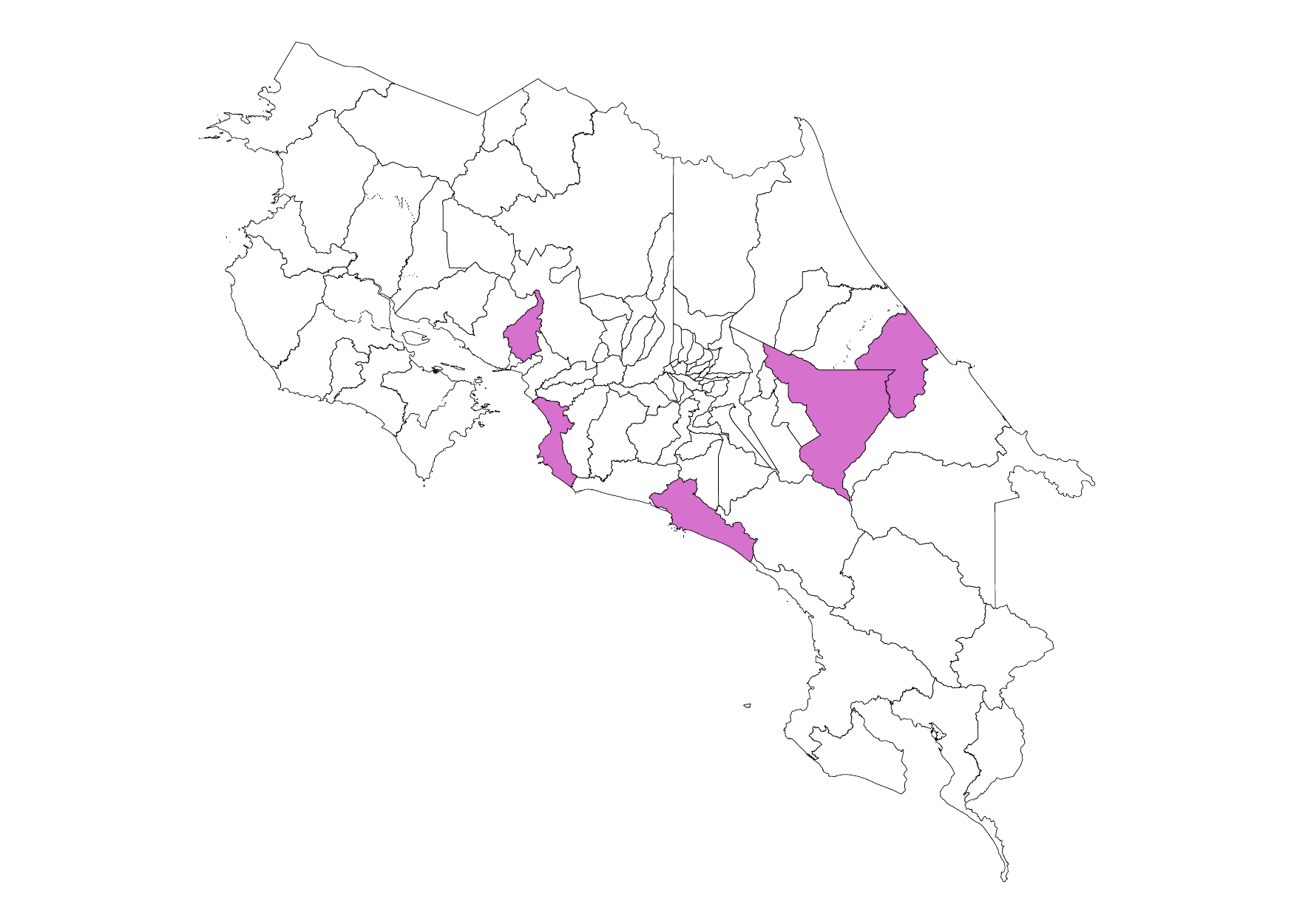}
\end{minipage}
&
\begin{minipage}{11.5cm}
\vspace{0.2cm}
There is only an area of joint significance between climate variables and dengue cases in Garabito and Montes de Oro in the period of 3-yr after 2010. TNA and dengue time series are in-phase with vanishing phase differences from 2011 to 2016, and El Ni\~no 3.4  has an advantage of approximately 9 months over dengue cases. For the remaining cantons, small areas of significance are observed in 1 and 2-yr periods with dengue time series ahead.
\end{minipage}\\
\end{tabular}
\end{table}

\subsection*{Local environmental variables and dengue cases}

\begin{figure}[H]
\centering
\includegraphics[scale=0.4]{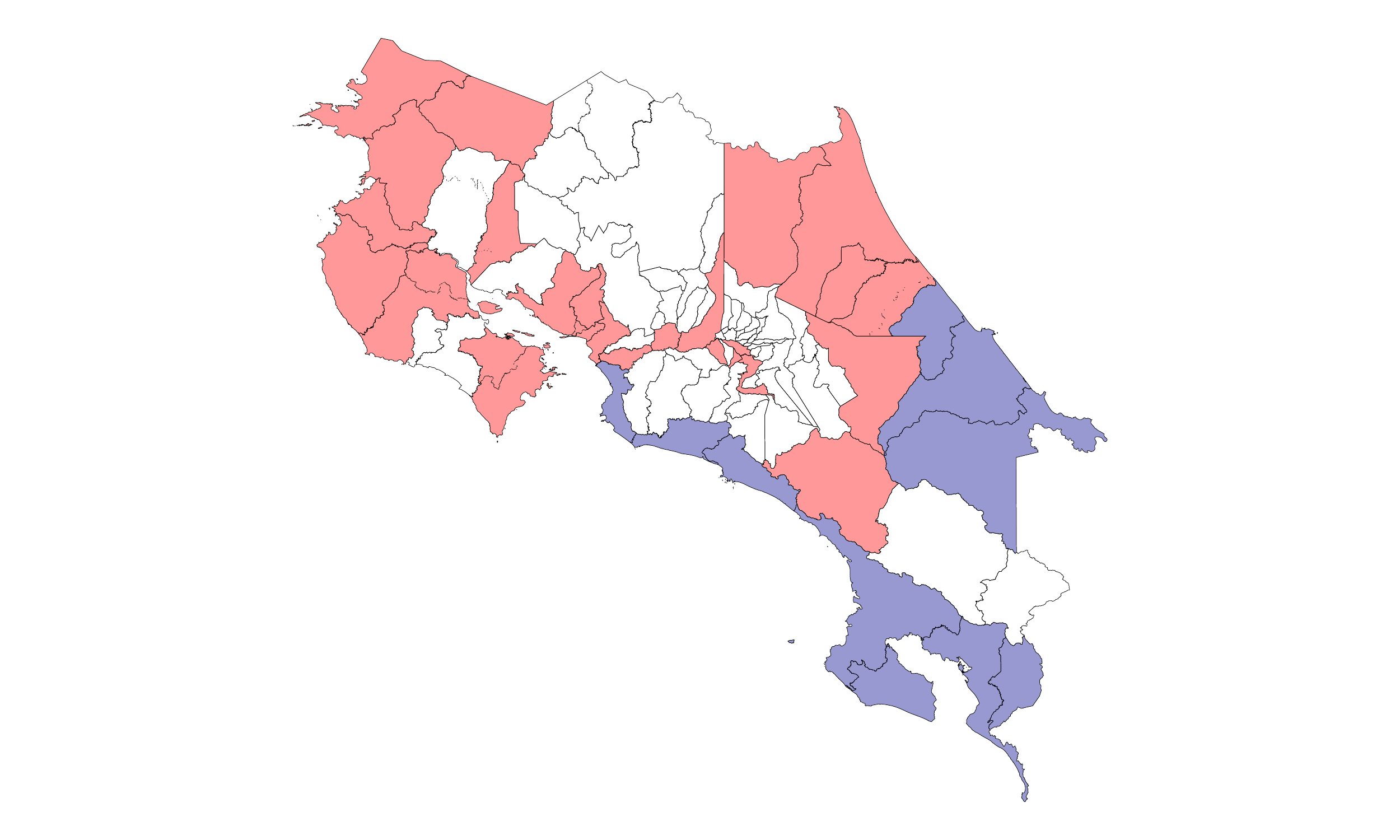}
\caption{{\bf Cantons where dengue cases co-varied with the  environmental variables.} Red cantons are those in which the annual dengue outbreaks was highly coherent with all environmental variables.}
\label{fig:EffecEnvir}
\end{figure}

Using coherence analysis to compare these time series in the frequency domain, we found that the annual dengue outbreaks were highly coherent with all environmental variables. In all cases, the phase difference shows an increase in dengue incidence after an increase in the local environment variables with a lag ranging between 0 to 3 months. The association was nonstationary, with disruption from around 2007 to 2010. Dengue incidence showed significant coherence with the four variables EVI, NDWI, ET, and precipitation except in Lim\'on, Matina, Talamanca, Corredores, Golfito, and Osa, located in the South Pacific and South Caribbean of the country, and Parrita, Garabito, and Quepos in the central Pacific (see Fig~\ref{fig:EffecEnvir}). The wavelet coherence for the last cantons showed areas of high significance between dengue, mainly with ET and precipitation in short periods. This finding is expected due to the regular seasonality observed on the Pacific coast and the country's Central region. The main differences observed between the clusters are the periods in which the synchronicity of the dengue cases and local environmental time series is observed.

\subsubsection*{Characteristics of clusters that include local environmental variables (LEV)}

\begin{table}[H]
\centering
\begin{tabular}{p{4cm}p{11.5cm}}
\textbf{Cluster 1}& \textbf{Alajuela, Montes de Oro, Orotina, Parrita,
Pococci, Santa Ana, Turrialba}\\
\begin{minipage}{4cm}
\includegraphics[width=1.2\textwidth]{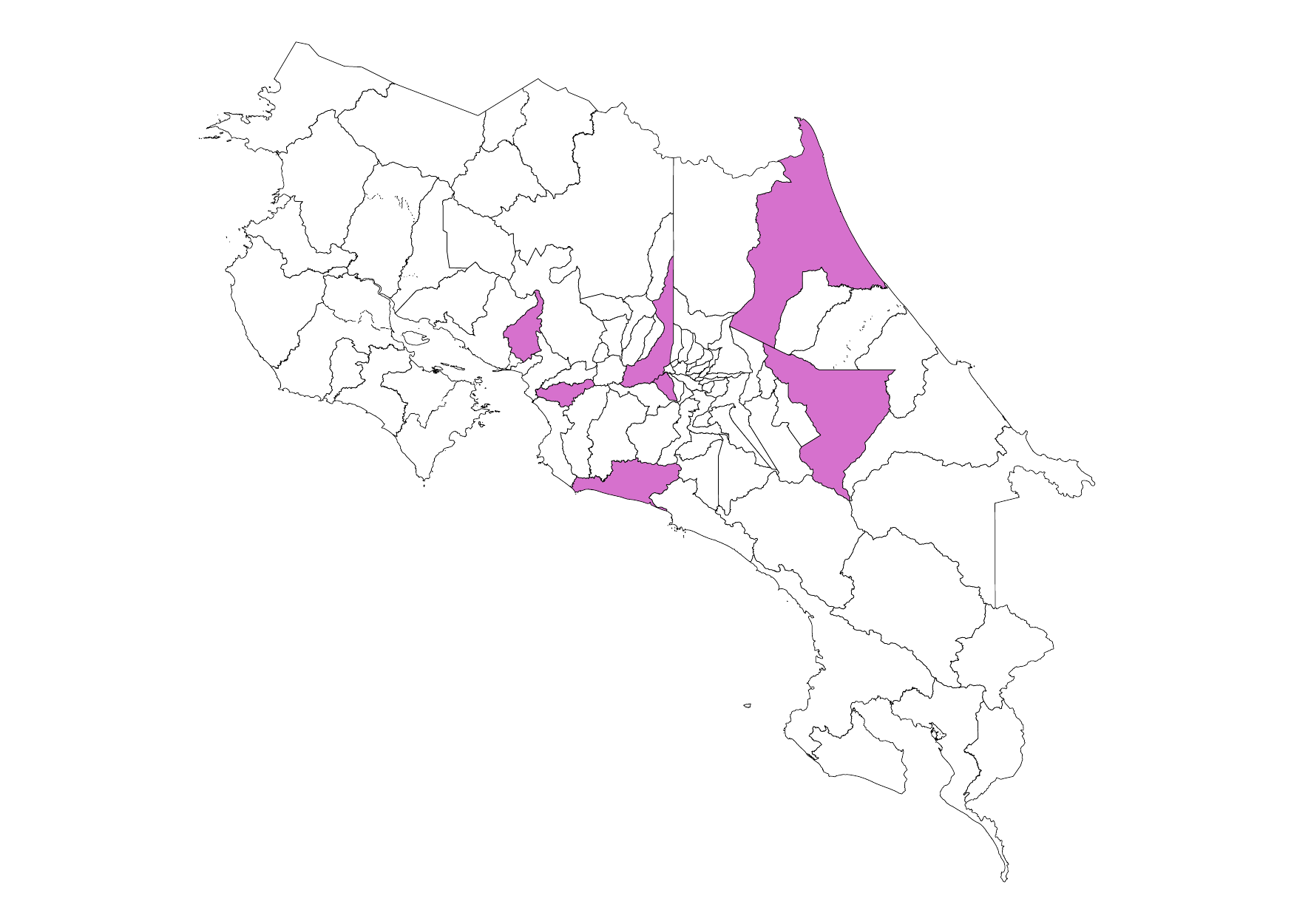}
\label{fig:ELVCluster1}
\end{minipage}
&
\begin{minipage}{11.5cm}
Annual dengue outbreaks showed significant coherence with the four environmental variables, except in Parrita and Pococí. In all cases, the leading time series were the environmental variables with 0 to 3 months ahead of dengue cases. Wavelet coherence in Parrita shows an area of joint significance only with precipitation before 2009 with the time series in phase. Wavelet coherence in Pococ\'i has a continuous area of joint significance with all variables. While for the rest of the cantons, a disruption is observed between the areas of significance around 2007 to 2010.\\
\end{minipage}\\
\end{tabular}
\end{table}

\begin{table}[H]
\centering
\begin{tabular}{p{4cm}p{11.5cm}}
\textbf{Cluster 2}& \textbf{Alajuelita, Lim\'on} \\
\begin{minipage}{4cm}
\includegraphics[width=1.2\textwidth]{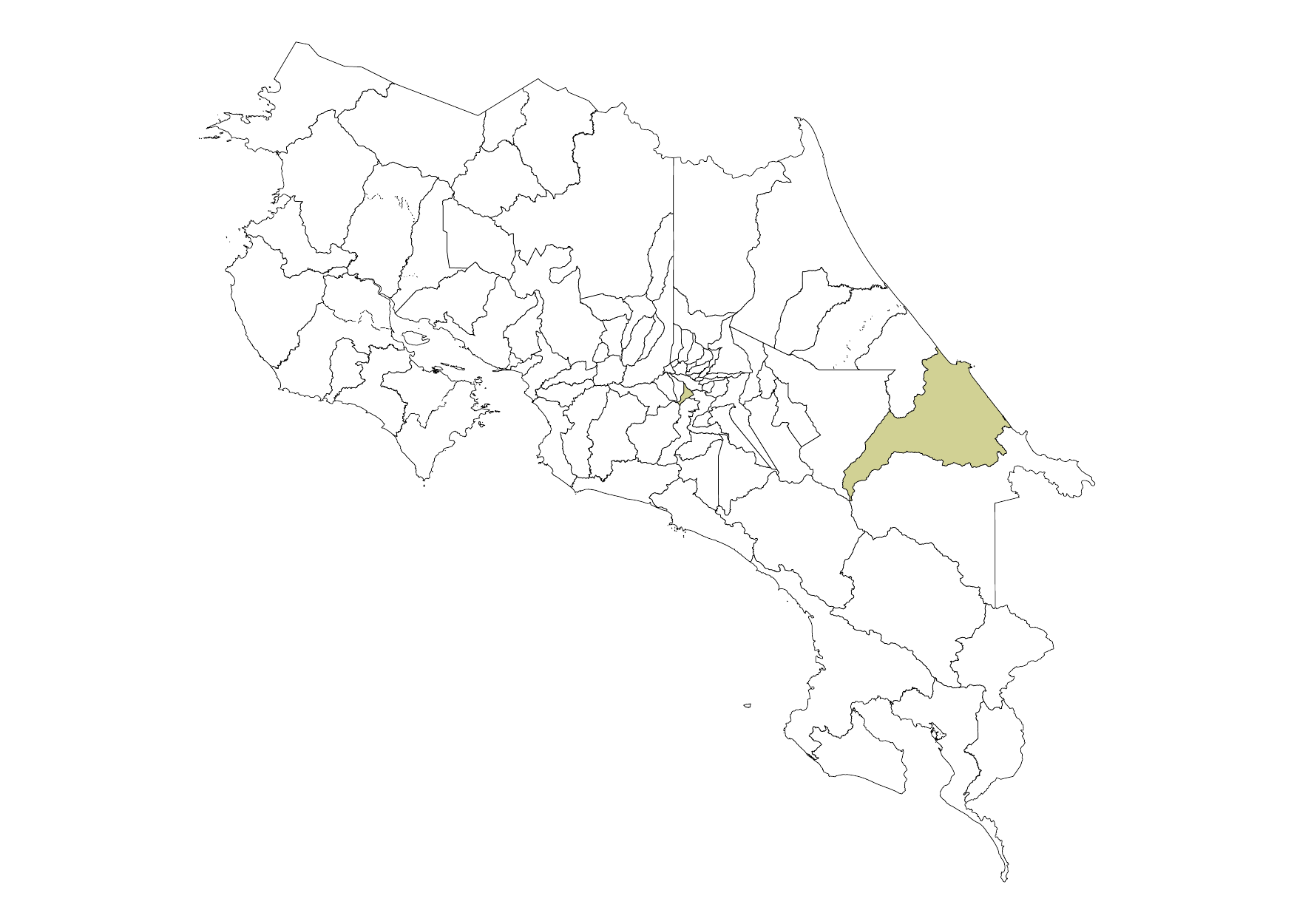}
\label{fig:ELVCluster2}
\end{minipage}
&	
\begin{minipage}{11.5cm}
Wavelet coherence in Alajuela shows two areas of joint significance with all variables in the 1-yr period with disruption around 2007 to 2010. The leading time series were the environmental variables with 0 to 3 months ahead of dengue cases. Wavelet coherence in Lim\'on shows an area of high significance only with ET in two short periods with an interruption between 2009-2010. The coherence with the other variables is not conclusive.
\end{minipage}\\
\end{tabular}
\end{table}

\begin{table}[H]
\centering
\begin{tabular}{p{4cm}p{11.5cm}}
\textbf{Cluster 3}& \textbf{Ca\~nas, Matina, Osa, Puntarenas} \\
\begin{minipage}{4cm}
\includegraphics[width=1.2\textwidth]{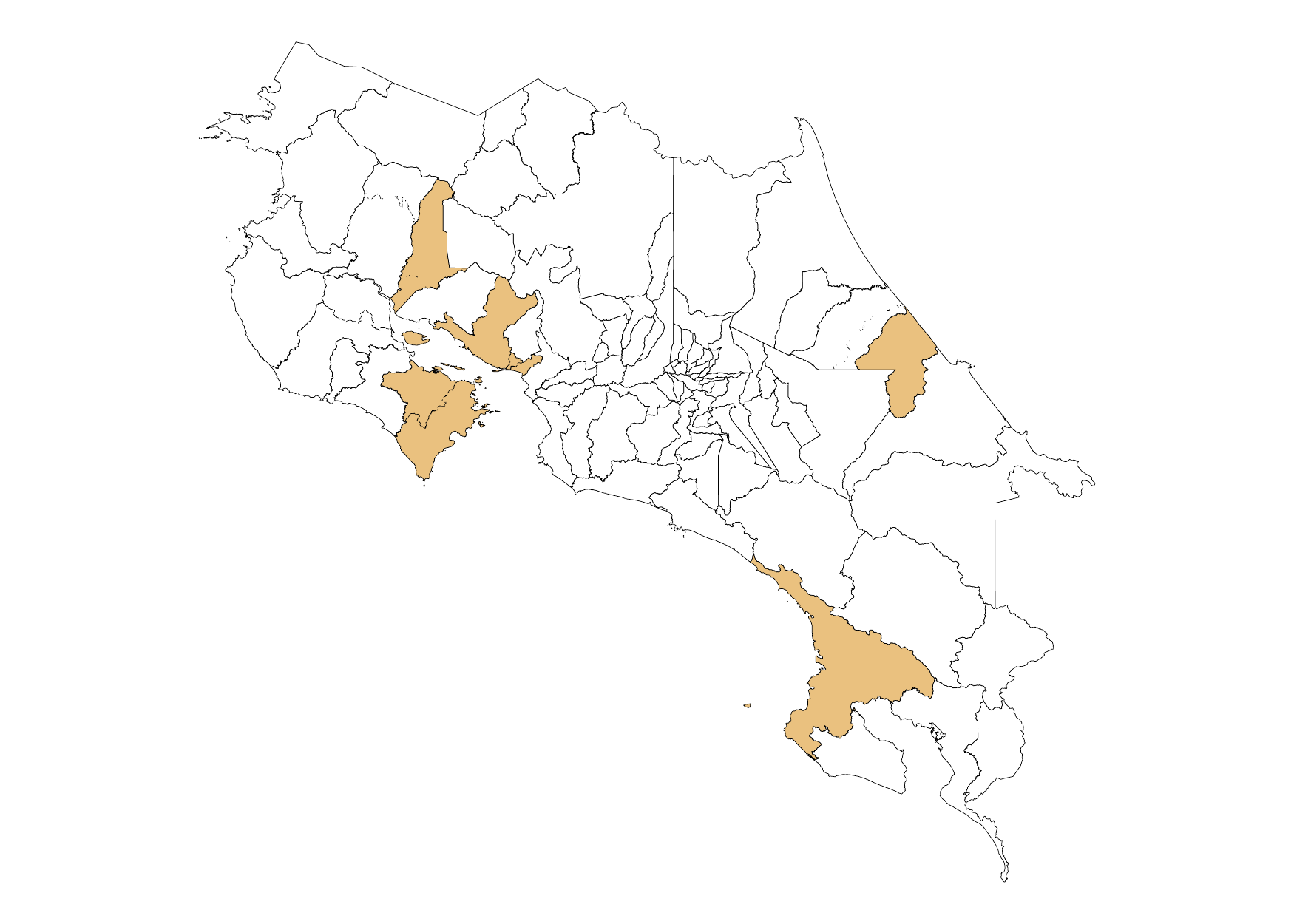}
\label{fig:ELVCluster3}
\end{minipage}
&
\begin{minipage}{11.5cm}
Wavelet coherence in Ca\~nas and Puntarenas show a correlation with the four vegetation variables in the 1 yr period (in Ca\~nas, there is a continuous area of joint significance over time, while in Puntarenas is interrupted between 2008-2010). Smaller areas of significance are observed between dengue cases in Osa and the vegetation variables, mainly after 2013. In all cases, environmental variables are the leading time series with a lag of fewer than 3 months. Wavelet coherence in Matina shows an area of high significance only with ET and EVI in 1-yr. The environmental time series go ahead by less than 3 months.\\
\end{minipage}\\
\end{tabular}
\end{table}

\begin{table}[H]
\centering
\begin{tabular}{p{4cm}p{11.5cm}}
\textbf{Cluster 4}& \textbf{Carrillo, Desamparados, Esparza, La Cruz, 
Perez Zeled\'on, San Jose, Santa Cruz, Sarapiqu\'i, Siquirres, Talamanca} \\
\begin{minipage}{4cm}
\includegraphics[width=1.2\textwidth]{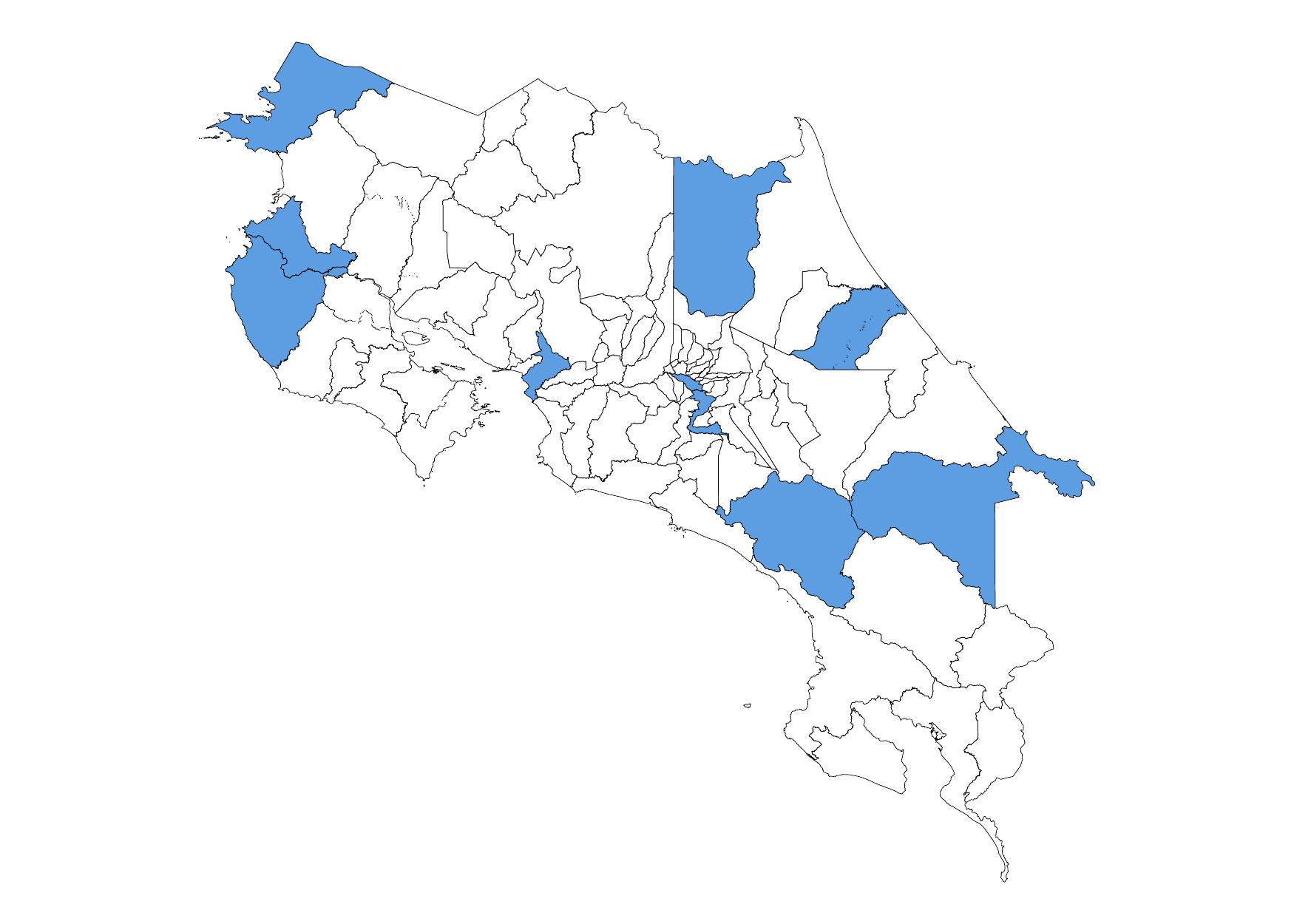}
\label{fig:ELVCluster4}
\end{minipage}
&	
\begin{minipage}{11.5cm}
Wavelet coherence shows areas of joint significance between dengue and the four vegetation variables in the 1-yr period with an interruption around 2009-2010, except for San Jos\'e and Talamanca. In all cases, the leading time series were the environmental variable by approximately 3 months ahead of dengue cases. There is no observed coherence between ET and dengue cases in San Jos\'e, and dengue cases in Talamanca showed significant coherence only with ET in a short period (around 2011-2015). The area of high significance is smaller in Perez Zeled\'on, Sarapiqu\'i, and Siquirres than in the other cantons in the cluster. Wavelet coherence in Talamanca shows an area of joint significance only with ET in a short period (2011-2015 approx)\\
\end{minipage}\\
\end{tabular}
\end{table}

\begin{table}[H]
\centering
\begin{tabular}{p{4cm}p{11.5cm}}
\textbf{Cluster 5}& \textbf{Corredores, Golfito, Liberia, Nicoya}\\
\begin{minipage}{4cm}
\includegraphics[width=1.2\textwidth]{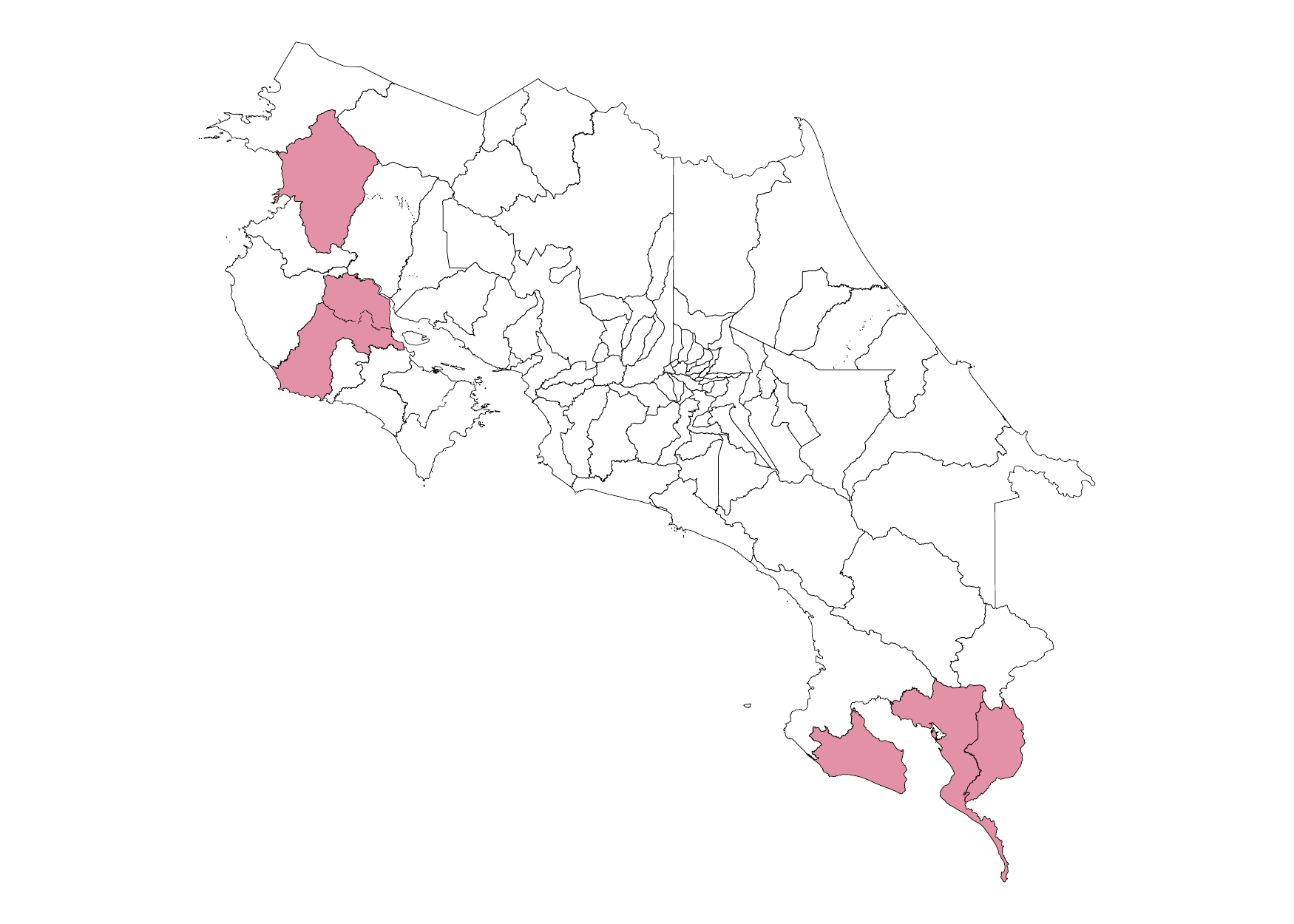}
\label{fig:ELVCluster5}
\end{minipage}
&
\begin{minipage}{12cm}
Wavelet coherence shows areas of joint significance between dengue cases in Liberia and Nicoya with the four vegetation variables in the 1-yr period with minor interruptions over time. Dengue cases in Corredores and Golfito showed significant coherence only with  ET and precipitation in the 1-yr period. Local environmental variables are the leading time series in all cases by approximately 3 months ahead of dengue cases.
\end{minipage}\\
\end{tabular}
\end{table}

\begin{table}[H]
\centering
\begin{tabular}{p{4cm}p{11.5cm}}
\textbf{Cluster 6}& \textbf{Garabito, Gu\'acimo} \\
\begin{minipage}{4cm}
\includegraphics[width=1.2\textwidth]{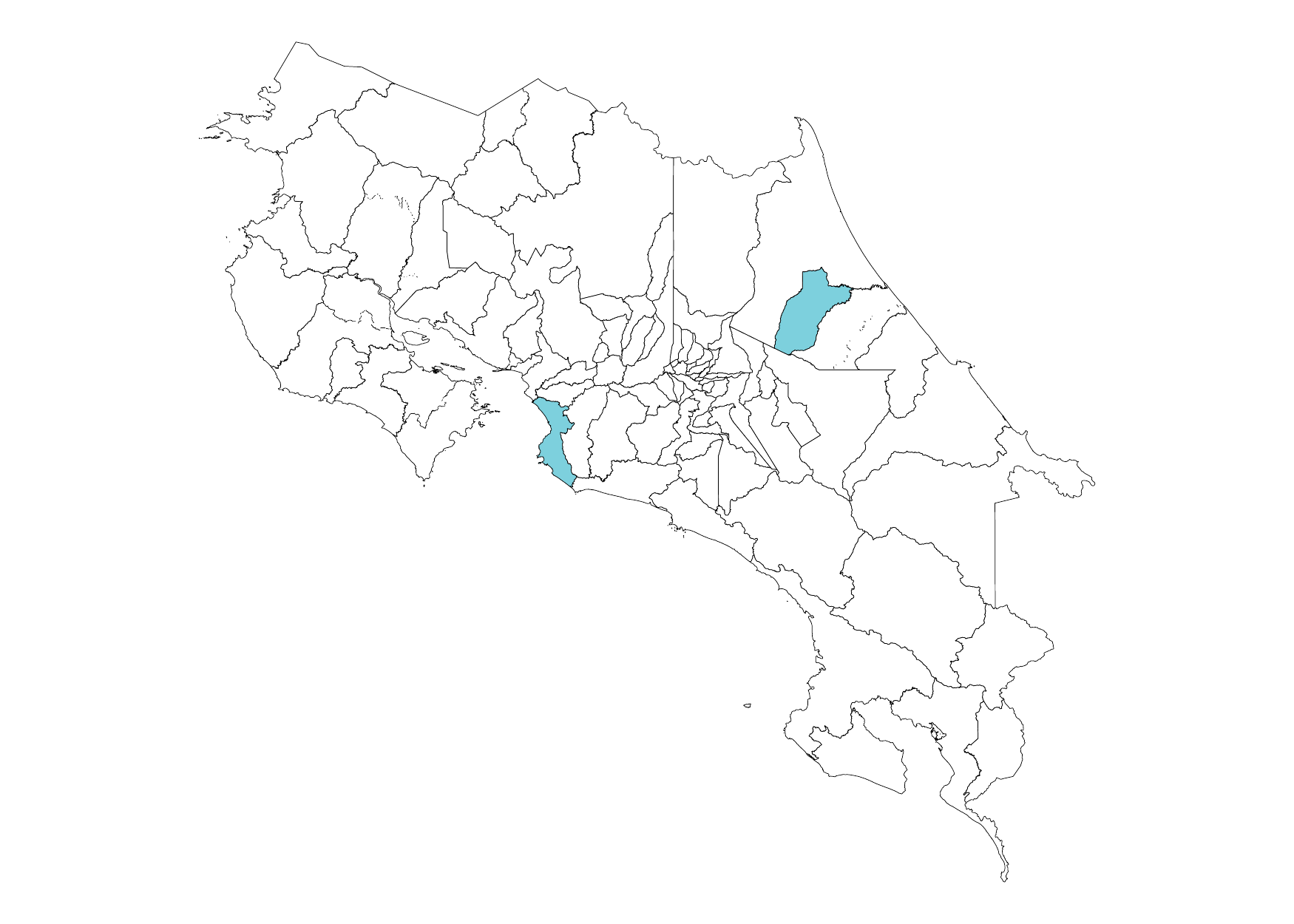}
\label{fig:ELVCluster6}
\end{minipage}
&	
\begin{minipage}{11.5cm}
Dengue time series in Gu\'acimo showed significant coherence with all variables after 2009 in the 1-yr period, with the environmental variables ahead of dengue cases by approximately 3 months. Dengue time series in Garabito showed significant coherence only with ET and precipitation before 2012 in the 1-yr period.\\
\end{minipage}\\
\end{tabular}
\end{table}

When creating the clusters, Atenas, Upala, and Quepos stayed isolated as the unique canton in their groups. Quepos is a touristic canton located in the central Pacific where the rain is abundant. The wavelet coherence shows an area of joint significance only between dengue cases and ET and precipitation for a short period, with the dengue time series the leading one. Upala is located in the North in a subregion where the climate is classified as rainy with monsoon influence (in essence, a tropical monsoon climate tends to have more precipitation and less pronounced dry seasons). The wavelet coherence shows a region of high joint significance between 2012 and 2015 (the worst dengue outbreak in Upala was registered in 2013) with all variables except EVI. The leading time series is dengue except in the coherence with ET, where the time series were in phase with a vanish phase difference. Atenas is located in the Central of the country. The wavelet coherence shows two areas of joint significance in the 1-yr period, with a disruption between 2007-2009. The time series are in-phase with environmental variables leading, except between 2009-2013, when the dengue time series started to be the leading time series. The lag change between 0-3 months.
\section*{Discussion}
\label{Discussion}

\subsection*{Climate variables}

Our results show that the 3-year dengue outbreaks were highly coherent with El Ni\~no 3.4 and TNA indicators, mainly in cantons located in the North, South Pacific coastal, and Center region of Costa Rica. When analyzing phase differences between the time series in these regions, synchrony between dengue cases and climatic variables was observed. The time series of El Ni\~no and TNA used to be ahead of dengue cases by nine and less than three months, respectively. In some cantons, TNA and dengue cases are almost in perfect sync. The synchronization of the time series began to be observed mainly after 2008. The periods in which areas of high significance become evident and the year in which the time series starts to be synchronized change between cantons and over time. Those timing periods and the series leading are the main differences between the clusters.

Our results also showed that annual dengue outbreaks were highly coherent with TNA in cantons located in the North Pacific and the Central region of the country, with the dengue and TNA time series synchronized for short periods with a vanishing phase difference. In some cantons, associations between climate and dengue cases were also observed in the 1, 1.5, and 2-year periods, with dengue time series leading. However, dengue cases influencing climate variables are a biologically unlikely relationship. It is more plausible that changes in El Ni\~no 3.4 and TNA increase subsequent dengue transmission~\cite{tipayamongkholgul2009effects}, given that the anomalies in the ocean temperatures affect local temperature and precipitation, which are directly associated with the virus ecology and transmission~\cite{kolivras2010changes,morin2013climate,tun477effects}.

Significant coherence between climate variables and dengue cases was observed mainly in the cantons in which there are periods of drought. There is no evidence of a contemporary statistical association between the SST anomalies in the two oceans. But, it has been found that seasonal precipitation is dependent upon the simultaneous interaction of both anomalies. The draught, which typifies the Pacific slope during years of warm phase ENSO, is less severe when the Caribbean is warmer, but the reduction in rainfalls may spread over the entire country~\cite{quesada2001effect}. The North Pacific belongs to the Pacific precipitation regime, known for a well-defined dry and rainy period~\cite{solanosf}. The geographical characteristics in the central and South Pacific generate a climate where the dry period is very favorable and short and the rainy intense~\cite{solanosf}. The years of intense droughts in those regions, except for 2001, have coincided with El Ni\~no years~\cite{de2008clima}. There has been observed that during El Ni\~no, there is a greater probability that the entire Pacific slope and the Central region of Costa Rica will experience dry to extremely dry conditions. At the same time, there is a greater probability of extreme rainy scenarios in the Caribbean~\cite{de2008clima}.

\subsection*{Environmental variables}

The annual dengue cycle was highly coherent with the four local environmental variables in all cantons except those located in the central Pacific, South Pacific, and South Caribbean. For those, dengue incidence showed a significant coherence with ET and precipitation over short periods. In those regions, the rain is abundant, and there is not a well-defined dry season~\cite{quesada2001effect} and the time series for the vegetation indices (EVI, NDWI) does not show an annual seasonality like that observed in the North Pacific and the Central valley of the country. Regarding synchronization, dengue and environmental time series were in phase with the environmental time series ahead of dengue cases by approximately 3 months.

Climate and environmental variables showed significant coherence with dengue epidemics across the geographically diverse regions of Costa Rica. However, spatial heterogeneity in their effects exists. Even when the association was established, it was not possible to provide those variables' causal effects on the dengue cases' dynamic. Epidemic data is typically noisy, complex, and non-stationary. Changes in the periodicity over time may also be to external factors or inherent characteristics of the disease. Alternative explanations for the multi-annual outbreaks of dengue epidemics have been proposed, such as partial cross-immunity among the four serotypes of the dengue virus~\cite{adams2006cross}. Moreover, sociological factors such as population structure, unplanned urbanization, and international transportation of infected people and mosquitoes may also affect dengue transmission~\cite{kuno1995review}. 

The complex interaction of biological, socioeconomic, environmental, and climatic factors creates a substantial spatiotemporal heterogeneity in dengue outbreak intensity. Understanding this heterogeneity in dengue transmission could improve epidemic prediction and disease control. There are some previous works in which predictive models have been proposed for the country using climatic and vegetation variables. Fuller et al.~\cite{fuller2009nino} proposed a simple structural model incorporating lagged SST and MODIS vegetation indices, explaining 83\% of the variance in the total weekly cases of DF/FHD in Costa Rica. However, the cantonal-level analysis conducted in this study highlights the spatial heterogeneity of the effect of climate and environmental factors on dengue incidence, which reveals that the effect of those variables on dengue transmission on a local scale might differ from global expectations. Thus, for the design of interventions and resource allocation, more localized predictions may be helpful. Other studies show that when incorporating the same climatic variables in all cantons to predict relative risk areas of dengue outbreak, the models projections may fail in some places~\cite{barboza2022assessing, vasquezclimate}, which reinforces the importance of understanding the local  correlations between dengue cases and external factor such as climate and socioeconomic drivers to improve local dengue predictions.

Many studies have been conducted in tropical and subtropical regions to elucidate the complex interactions between climate variables and dengue transmission~\cite{prabodanie2020coherence,ehelepola2015study, cuong2016quantifying,thai2010dengue,johansson2009multiyear,cazelles2005nonstationary,jury2008climate}. However, this study is unique in the localized analysis that takes into account  variables such as TNA, EVI, NDWI, evapotranspiration, and precipitation. Wavelet time serires analysis allows a retrospective study to characterize outbreaks over time, which provides important guidelines for future modeling approaches in which explicit mechanisms can be incorporated. However, as climate factors are not the only predictors influencing the rise in dengue infection, future studies are needed to include other factors unique to this area, such as the predominant circulating dengue viruses, anthropogenic factors, and herd immunity, to name a few.



\section*{Acknowledgments}
This research was supported by a Seed Grant for International Activities from Global Affairs and the School of Medicine at the University of California, Davis.

\section*{Additional information}

\textbf{Codes}: All the codes and files necessary to reproduce the results presented in this document will be available in the GitHub repository, \url{https://github.com/yurygarcia26/Wavelets_Costa_Rica.git}, after publication.\\

\nolinenumbers

%
%
%

\end{document}